\documentclass[preprint,12pt]{elsarticle}

\usepackage{amsmath}
\usepackage{amssymb}
\usepackage[T1]{fontenc}
\usepackage{xcolor}
\usepackage[pagewise]{lineno}
\usepackage{changes}
\renewcommand{\added}[1]{{\color{black}#1}}
\renewcommand{\deleted}[1]{}

\begin{document}

\begin{frontmatter}



\title{Molecular dynamics simulations of metal-electrolyte interfaces under potential control}


\author[inst1]{Linn\'{e}a Andersson}

\affiliation[inst1]{organization={Department of Chemistry-\AA{}ngstr\"{o}m Laboratory, Uppsala University},addressline={L\"{a}gerhyddsv\"{a}gen 1, BOX 538}, city={Uppsala}, postcode={75121}, country={Sweden}}


\author[inst1]{Chao Zhang}
\ead{chao.zhang@kemi.uu.se}
\begin{abstract}
The interfaces between metal electrodes and liquid electrolytes are prototypical in electrochemistry.  That is why it is crucial to have a molecular and dynamical understating of such interfaces for both electrical properties and chemical reactivities under potential control. In this short review, we will categorize different schemes for modelling electrified metal-electrolyte interfaces used in molecular dynamics simulations. Our focus is on the similarities between seemingly different methods and their conceptual connections in terms of relevant electrochemical quantities. Therefore, it can be used as a guideline for developing new methods and building modularized computational protocols for simulating electrified interfaces.  
\end{abstract}



\begin{keyword}
 molecular dynamics simulations \sep density functional theory \sep electrochemical interfaces \sep constant potential \sep finite field methods
\end{keyword}

\end{frontmatter}


\section{Introduction}

The metal-electrolyte interface plays a key role in electrochemical energy storage and conversion. In general, there are two kinds of \added{\textit{idealized}} electrodes: ideal polarizable and ideal non-polarizable. The former as exemplified by the Hg electrode involves only capacitive charging and zero exchange current (density); the latter, taking the example of the Ag/AgCl electrode, allows free passing of faradic current with zero charge transfer resistance~\cite{Schmickler:2010th}. Different from the meaning of dipole-moment density in physical chemistry, polarization here refers to the applied potential. In other words, an ideal polarizable electrode can sustain an applied potential and form an electric double layer. Therefore, a realistic \added{computational} model of the metal-electrolyte interface under electrochemical conditions should include the effects of an external potential.

 Despite that the grand canonical (GC) formulation of density functional theory (DFT) was introduced right after the birth of canonical DFT in the mid 60s~\cite{Mermin1965}, its implementation for a slab system which resembles an metal-electrolyte interface in electronic structure calculation came out much later. Notably, Lovozoi et al.~\cite{lozovoi} discussed strategies of applying GC-DFT to metal slab systems under periodic boundary conditions (PBCs) already in 2001. In particular, they realized that PBCs convoluted the electrostatic potential of systems \added{\textit{with a net charge}} because of the homogeneous compensating background and that an electrostatic correction must be applied to restore the physical one. 

 Parallel to the development of GC-DFT approaches, attempts were made to use charge transfer from ions to the metal interface to mimic the electrified interfaces with canonical DFT. Early examples were given by Skúlason et al. \cite{skulason} and Rossmeisl et al.~\cite{2008.Rossmeisl} around 2007, where different numbers of hydrogen atoms were added at the the Pt-water interface, to study the system at different electrode potentials. In these setups, the hydrogen atoms turn to solvated protons in the water bilayer and release electrons to the metal surface, which provides a static model of the electric double layer. 

 Since metal-electrode interfaces involve both electronic and ionic responses (to an external potential), a dynamical picture of such interfaces beyond a static DFT description is clearly needed. The pioneering work in this direction was done by Price and Halley in 1995~\cite{Price:1995fs}. They carried out the first Car-Parrinello-type molecular dynamics (MD) simulations of metal-water slab systems with an applied potential difference between the slabs (of 1.36 V), plane-wave basis set and local pseudopotential. 

These seminal works that developed independently have influenced later theoretical works that try to describe structural, dynamical and electronic properties of electrified interfaces between metal electrode and electrolyte solution on an equal footing. Therefore, an up-to-date view of this topic is desirable. On this note, we will focus on works based on DFTMD approaches and highlight the recent method developments in this field. Thus, discussions on their applications in electrocatalysis and electrochemical energy storage, can be found in excellent reviews elsewhere~\cite{gros_modelling_2019, 2019.Huang, 2021.Lei2j, Scalfi:2021et, 10.1002/wcms.1499, 2022.Sundararaman, 2022.Morgan}. Furthermore, we will only include methods based on equilibrium calculations, therefore, non-equilibrium methods such as Green's functions~\cite{10.1039/c7sc02208e} are not discussed here. 

In the following, we will first introduce necessary concepts in electrochemistry, which will help us to understand the motivation and the strategies behind different types of methods. Then, we will sort seemingly different methods into three categories and discuss examples in each type of method. Finally, a summary and outlook for future developments is also given. 

\section{Theoretical background}
\label{sec:sample1}
\subsection{Fermi level and electrode potential}

\begin{figure}[ht]
\begin{center}
\includegraphics[width=0.8\linewidth]{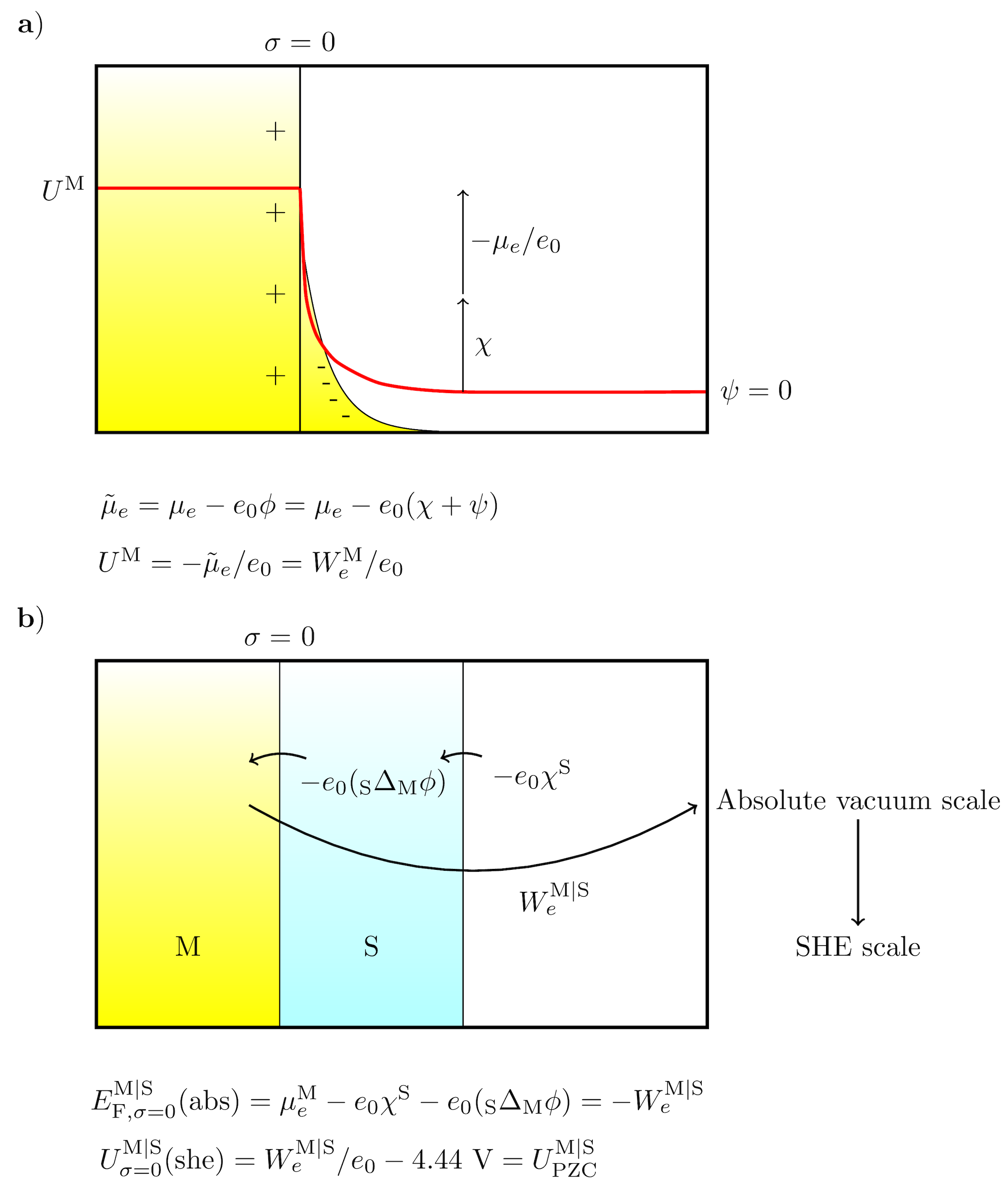}
\end{center}
\caption{\added{\textbf{Metal interfaces with vacuum and solution.} a) Metal-vacuum interface. The electrochemical potential $\tilde{\mu}_e$ consists of the chemical potential $\tilde{\mu}$ and a contribution from the Galvani potential $\phi$. The Galvani potential is composed of the Volta potential $\psi$ due to net surface charge and the surface (dipole) potential $\chi$. With zero surface charge, $\psi = 0$. b) Metal-solution-vacuum interfaces. The work function $W_e^\textrm{M|S}$ is defined as the energy required to take an electron from the metal through the solution and into vacuum. The potential of zero charge $U^ \textrm{M|S}_\textrm{pzc}$ can be calculated from this work function if the reference is changed from vacuum to standard hydrogen electrode (SHE). } \label{metal_if}}
\end{figure}

\added{The work function of a metal surface $W_e^\textrm{M}$ is defined as the energy required to take an electron from it into vacuum.} For bulk metal, it is well-known that the opposite of the metal work function equals to the Fermi level of the metal \added{$E_\textrm{F}^\textrm{M}$. The Fermi level is equal to the electrochemical potential $\tilde{\mu}^\textrm{M}_e$, which can be partitioned into the chemical potential $\mu^\textrm{M}_e$ and the contribution from the Galvani potential $\phi^\textrm{M}$ (see Figure \ref{metal_if}a for illustration)}:
\begin{equation}
    \label{fermi_gas}
    E_\textrm{F}^\textrm{M} = \tilde{\mu}^\textrm{M}_e =\mu^\textrm{M}_e - \phi^\textrm{M} e_0 =-W_e^\textrm{M} 
\end{equation}
Then, the corresponding thermodynamic electrode potential will be written as:
\begin{equation}
    \label{pot_gas}
    U^\textrm{M}  = W_e^\textrm{M}/e_0 
\end{equation}
where the electron charge is defined as $-e_0$.

Similarly, the thermodynamic electrode potential of metal in solution equals to the work function of metal in solution \added{(see Figure \ref{metal_if}b)}. 
\begin{equation}
    \label{pot_sol_abs}
    U^\textrm{M|S}_{\sigma = 0}(\textrm{abs}) = W_e^\textrm{M|S}/e_0
\end{equation}

This is a key conclusion from a series of works from Trasatti and elaborated in Section 2.4 in Ref.~\cite{Cheng:2012cj}. In the expression above, we emphasize that the work function is well-defined when the surface charge $\sigma$ is zero as by definition it does not include any contributions from the \added{Volta potential $\psi$ (see Figure \ref{metal_if}a and its caption)}. 
Subsequently, it means the corresponding Fermi level of the metal electrode in solution can be expressed as:
\begin{equation}
    \label{fermi}
    E_{\textrm{F}, \sigma = 0}^\textrm{M|S} = - W_e^\textrm{M|S} 
\end{equation}

For the electrified metal-electrolyte interface, the double-layer potential $\phi_\textrm{EDL}$ will be built up and this leads to the expression of the Fermi level as:
\begin{equation}
    \label{fermi_solution}
    E_\textrm{F}^\textrm{M|S} =  E_{\textrm{F}, \sigma = 0}^\textrm{M|S} - e_0\phi_\textrm{EDL} 
\end{equation}

In addition, by definition, the double-layer potential is also equal to the change in the Galvani potential difference \added{$_\textrm{S}\Delta_\textrm{M}\phi = \phi^\textrm{M} - \phi^\textrm{S}$ between metal (M) and solution (S)}, i.e. 
\begin{equation}
    \label{fermi_EDL}
    \phi_\textrm{EDL} = _\textrm{S}\Delta_\textrm{M}\phi - _\textrm{S}\Delta_\textrm{M}\phi^{\sigma=0}
\end{equation}

As noted by Trasatti, despite ``that the experimentally measured potential has nothing to do with the actual electric potential drop across the interface''~\cite{Gerischer1977-st},  they change by the same amount when the metal electrode is polarized. Thus, one can adjust the change in the Fermi level of metal-electrolyte interfaces in order to control the double-layer potential. 
\begin{equation}
    \label{fermi_inner}
    \Delta E_\textrm{F}^{\textrm{M|S}} = -e_0\Delta(_\textrm{S}\Delta_\textrm{M}\phi) = -e_0\phi_\textrm{EDL}
\end{equation}

As shown in Section~\ref{examples}, this idea is exploited in grand canonical DFT or constant Fermi DFTMD, in which the total number of electrons in the system is varied in order to control the \emph{change} in the Fermi level. 

\subsection{Charge transfer and Volta potential}
 The alternative way to realize the ``constant potential" is through charge transfer within the whole system itself (in contrast to the external reservoir) . In the following, we will use two examples to illustrate this point. 

\begin{figure}[ht]
\begin{center}
\includegraphics[width=0.8\linewidth]{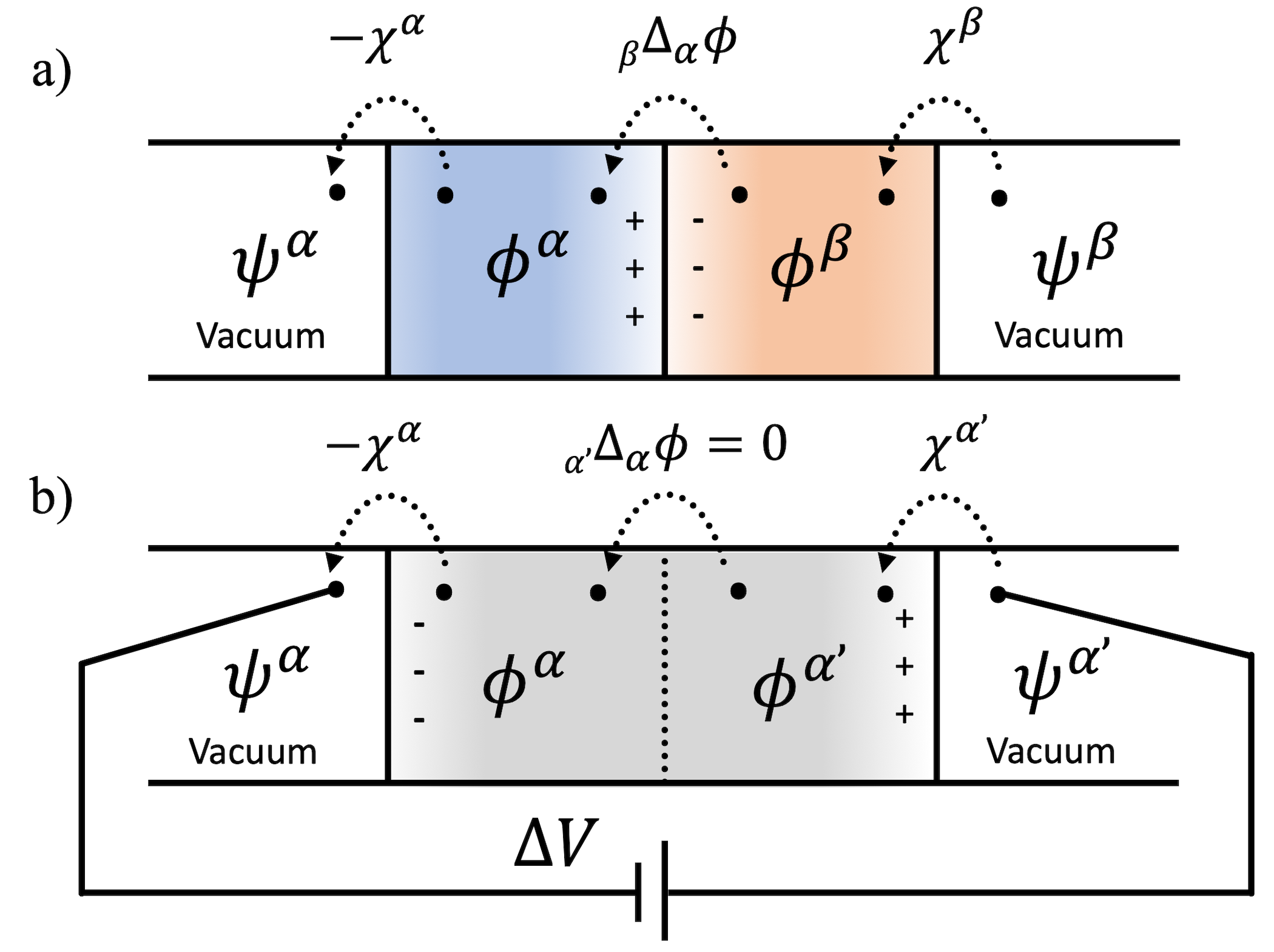}
\end{center}
\caption{\textbf{Generating the Volta potential difference due to charge transfer.} a) Two metals of different species in contact in which the metal $\alpha$ has a smaller work function. The Volta potential difference equals to the opposite of the work function difference. b) Two metals of the same species in contact under an applied voltage $\Delta V$. The Volta potential difference equals to the applied voltage. The definition of symbols can be found in the Text.  \label{Volta_CT}}
\end{figure}

Supposing that we have two metal species $\alpha$ and $\beta$. When they are separated apart, the electrochemical potential or the Fermi level of each species is:
\begin{eqnarray}
\label{twometals_seperate}
    \tilde{\mu}^\alpha_e &=& \mu^\alpha_e - e_0\chi^\alpha \\
    \tilde{\mu}^\beta_e &=& \mu^\beta_e - e_0\chi^\beta
\end{eqnarray}
where $\mu$ is the chemical potential and $\chi$ is the surface potential \added{(see Figure \ref{metal_if}a and its caption)}. 

Then, we put them in contact, as shown in Fig.~\ref{Volta_CT}a. Supposing $\tilde{\mu}_\beta$ is more negative as compared to $\tilde{\mu}_\alpha$, then a charge transfer will take place and the two metal species will become electrified. Now the electrochemical potential of each species is:
\begin{eqnarray}
\label{twometals_contact}
    \tilde{\mu}'^\alpha_e &=& \mu^\alpha_e - e_0\chi^\alpha - e_0\psi^\alpha \\
    \tilde{\mu}'^\beta_e &=& \mu^\beta_e - e_0\chi^\beta - e_0\psi^\beta 
\end{eqnarray}
where the Volta potential $\psi$ is no longer zero. 

The electrochemical potential of two metal species in contact must be the same. This means the chemical potential, the surface potential and the Volta potential have the following relation:
\begin{eqnarray}
 e_0(_\beta\Delta_\alpha\psi) &=& e_0(\psi^\alpha - \psi^\beta)  \\
 &=& \mu^\alpha_e - e_0\chi^\alpha - (\mu^\beta - e_0\chi^\beta) \\
 &=& W_e^\beta - W_e^\alpha
\end{eqnarray}
where we have used the relation between the work function, the chemical potential and the surface potential. It is worth noting that in this conceptualization, the change in the surface potential $\chi$ upon contact is not considered. 

The significance of the above equations is that the system of two metal species in contact is actually under a constant potential, since the work function of each metal species is an intrinsic material's property. This means one could use a counter-electrode or counter-ion which has a different work function to control the Volta potential difference. 

Along the same line, the follow-up example is two metals of the same species in contact under an external voltage $\Delta V$ (Fig.~\ref{Volta_CT}b). Since two metals must be in equilibrium, this implies:
\begin{equation}
    e_0\Delta V = e_0(_{\alpha}\Delta_{\alpha'}\psi) = W_e^{\alpha} - W_e^{\alpha'} 
\end{equation}
In other words, charge transfer happens between two metals of the same species under an external voltage $\Delta V$. In fact, since no field can be sustained within the same species of a metal, this means the two surfaces of the same metal are electrified. 

Comparing two scenarios shown in Fig.~\ref{Volta_CT}, we can see both systems are under potential control without varying the number of electrons. This means the whole system is canonical while each of two metals in contact is grand canonical. As shown in Section~\ref{examples}, this can be realized through either counter-ion/pseudo-atom methods or finite-field methods. 

\section{Potential control with half cell and full cell models} \label{examples}

\subsection{Type I: grand canonical DFT and constant Fermi DFTMD}
In grand canonical DFT the number of electrons in the system is allowed to vary over time, which can be used to charge an electrode. The electrode Fermi-level is then forced to become equal to $\tilde \mu_e^{\text{target}}$. The change in Fermi level corresponds to a build up the double layer potential as shown in Equation \ref{fermi_inner}. During constant Fermi-level DFTMD, the electrochemical potential is set to a target value during SCF in practice. This calculation may be difficult to converge which is why multiple canonical DFT calculations were often used instead to approximate the grand canonical ensemble~\cite{2018.Kastlunger}. Another approach by Bonnet et al. to achieve this is to devise a potentiostat scheme where the calculation of the number of electrons is separated from the optimization of the electronic states \cite{Bonnet12}. The potential energy of the system in contact with an electron resevoir at $\tilde \mu_e^{\text{target}}$ is 

\begin{equation}
    E_{\text{tot}} = E(R,n_e) + \tilde \mu_e^{\text{target}} n_e
\end{equation}
and the derivative with respect to the number of electrons can be interpreted as a fictitious force:
\begin{equation}
    F_{n_e} = -\frac{\partial E_{\text{tot}}}{\partial n_e} = \tilde \mu_e^\textrm{M} - \tilde \mu_e^{\text{target}}
\end{equation}
Now we can let the electronic charge be a dynamical variable governed by 
\begin{equation}
    \dot n_e = \frac{p_{n_e}}{m_{n_e}}, \quad \dot P_{n_e} = F_{n_e} = \tilde \mu_e^\textrm{M} - \tilde \mu_e^{\text{target}}
\end{equation}
where $p_{n_e}$ and $m_{n_e}$ are fictitious momentum and mass. The fluctuation of the number of electrons about $\langle n_e \rangle $ then allows for sampling the grand canonical ensemble \added{to make $\langle \tilde \mu_e^\textrm{M} \rangle = \mu_e^{\text{target}}$}.

This method was used by Bouzid and Pasquarello \cite{Bouzid:2018bs} to study the charged Pt(111)-water interface. With the addition of a hydronium ion to the system, the electrode potential can be aligned to the standard hydrogen electrode (SHE) by considering hydrogen adsorbed to Pt as an intermediate step in the SHE reaction, similar to the proton insertion method that developed for doing the band alignment at the semiconductor-water interfaces~\cite{Cheng:2010gfa}.  This allowed them to relate properties of the interface, such as double layer capacitance, to experiment. 

The setup for constant Fermi-level DFT-MD is a half cell. The electrode charge then has to be compensated somehow to make the overall system charge neutral under PBCs.  Bonnet et al. used an electronic screening medium (ESM) whereas Bouzid and Pasquarello made use of a homogeneous background charge. The homogenous background charge results in spurious interactions and corrections to the energy and potential have to be included ~\cite{FilholNeurock}. Therefore, the key difference between different methods in the Type I category is about how the electrolyte model (therefore counter-charge) is designed and implemented. Indeed, various implicit solvation models have been develope for this purpose~\cite{Shankar15, Ringe17, Hormann19, 2021.Bhandari}. 

It is worth noting that simulations of the metal-electrolyte interface with GC-DFT can be done without explicitly referencing the Fermi-level. This point was illustrated by the constant inner potential (CIP) DFT method from Melander et al.~\cite{Melander2021}. If the inner potential in bulk solution is set to zero for both neutral and charged interfaces, $\phi_\textrm{S}$ = 0, then following Equation \ref{fermi_EDL} the double layer potential is 
\begin{equation}
    \phi_\textrm{EDL} = \phi^\textrm{M} - \phi^\textrm{M}_{\sigma = 0}.
\end{equation}
$\phi^\textrm{M}$ can be controlled by setting a constraint on the average potential in the bulk electrode. 

\begin{figure}[ht]
\begin{center}
\includegraphics[width=0.8\linewidth]{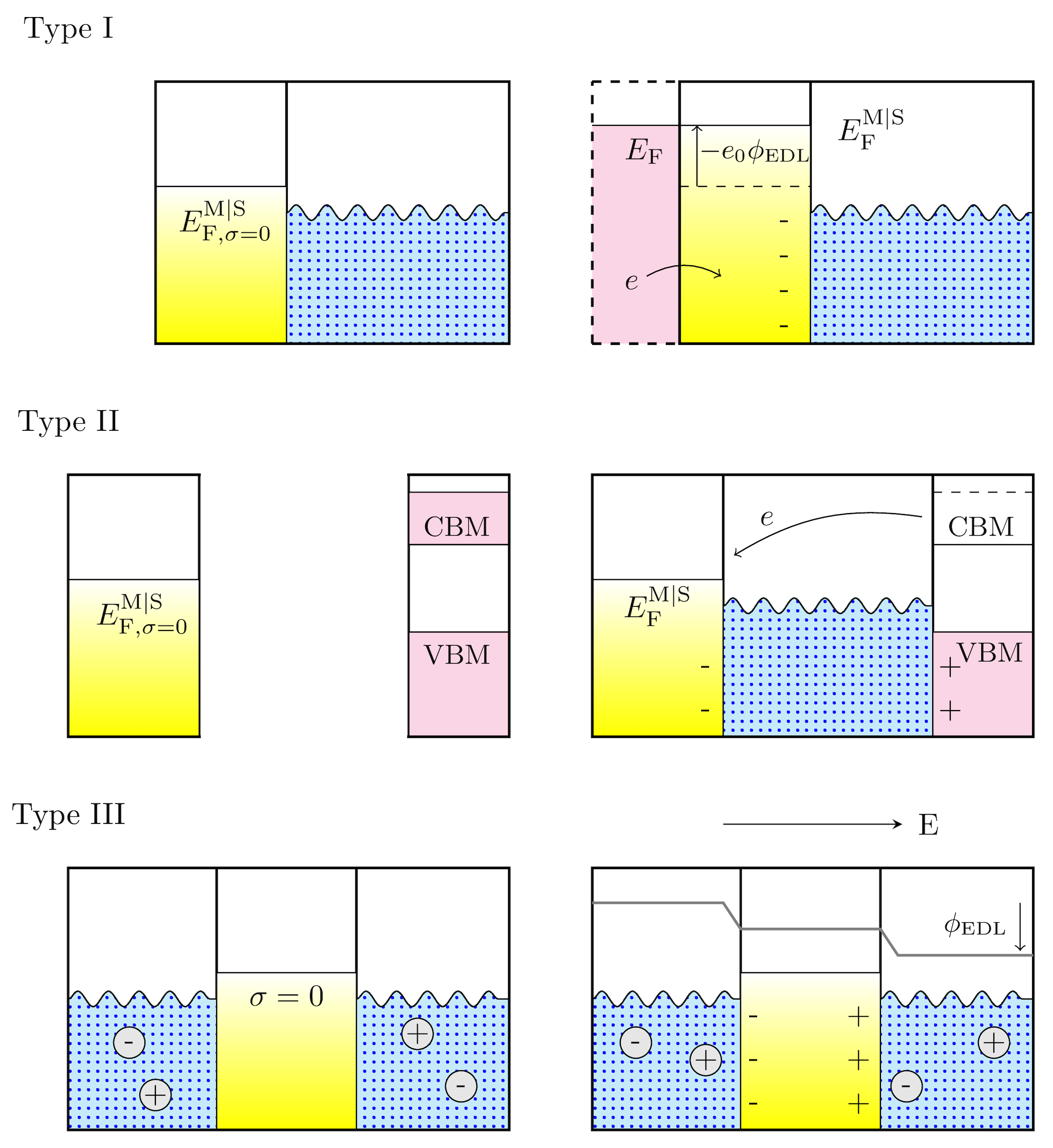}
\end{center}
\caption{\textbf{Three types of methods to impose the potential control for molecular dynamics simulations of metal-electrolyte interfaces.} Type I: grand canonical methods; Type II: Pseuduo-atom methods; Type III: Finite-field methods.  The dashed region is not part of the simulation box.\label{3_scheme}}
\end{figure}

\subsection{Type II: counter-ion/pseudo-atom methods}
In a full cell setup with two electrodes of the same metal, constant Fermi-level DFT cannot be used to induce a potential bias since the system only has one Fermi level. From the example in Figure \ref{examples}a) however, a system with two different metals in contact will have a potential bias across the cell stemming from the charge transfer induced by the equalization of their Fermi levels. A similar situation can also happen when a metal is put into contact with a charged semiconductor/insulator. If the Fermi level of the metal is in the band gap of the charged semiconductor/insulator, then any excess charge is going to transfer to the metal. 

Surendralal et al. \cite{Surendralal:2018fh} exploited this idea to control the potential of the anodic Mg/water interface using doped (pseudoatom) Ne as a counter-electrode. Ne was chosen for its large band gap and its valence band maximum/conduction band minimum relative to water. To generate a surplus or shortage of electrons in the system, the number of valence electrons and proton charge of each of the $n_\textrm{Ne}$ Ne atoms is changed by a fraction $q/n_\textrm{Ne}$. This way the total system stays charge neutral while each electrode obtains a charge $q$. To maintain a constant potential in the cell the counter-electrode charge can be altered during the dynamics. 
Similarly, in the work by Le et al. \cite{Le2019} the Pt electrode is charged by adding Na/F atoms close to its surface. Khatib et al. \cite{KHATIB2021138875} used the same strategy in their ion imbalance method but with Na/Cl atoms instead. In these setups, the electrode charge is constant (corresponding to the ion charge) while the double layer potential $\phi_\textrm{EDL}$ fluctuates. Although the double-layer can be modelled effectively in this manner, it is worth noting that the ion distribution is not fully in equilibrium as there is no real driving force to keep these ions next to the surfaces.

\begin{table}[h]
    \centering
    \begin{tabular} {c c c c c c c c}
       \hline
       Ref.  & Type & Cell & Electrolyte & Charge neutrality & SHE & PBC \\
       \hline 
       \cite{Bonnet12}  &  I & Half& None & ESM & No & 2D\\
       \cite{Bouzid:2018bs}  & I & Half & Water + H$_3$O$^+$ & Background charge & Yes & 3D\\
       \cite{Surendralal:2018fh}  &  II & Full & Water & Pseudo-atom & No & 3D\\
       \cite{Le2019}  & II & Half & Water + ions & Counter-ion & Yes & 3D \\
        \cite{Dufils:2019bk}  & III & Full & Water + ions  & Constraint & No & 3D \\
        \hline
    \end{tabular}
    \caption{\textbf{Examples of molecular dynamics simulations of metal-electrolyte systems under potential control.} }
    \label{tab:my_label}
\end{table}

\subsection{Type III:  finite-field methods}

Recently, Dufils et al. \cite{Dufils:2019bk} showed that constant potential simulation with the Siepmann-Sprik (SS) model which is commonly used in together with 2D PBCs can be also realized with 3D PBCs and the finite-field methods~\cite{Zhang:2020ks}. 

The basis of the SS model~\cite{siepmann_1995, reed_2007} is to allow the electrode charges to fluctuate in response to the external potential, which mimics the physical process of an ideal polarizable electrode. Although the original model was built for simulating metal-water interfaces under potential bias, its applications to electrochemical interfaces with electrolyte solution are straightforward. 

In the SS model, each response charge of the electrode atoms follows a Gaussian distribution of magnitude $c_i$ centered on the position of the electrode atom $\textbf{R}_i$
\begin{equation}
    \rho_i(\mathbf{r})=c_i\left(\frac{\zeta_i}{\pi}\right)^{3/2}\exp^{-\zeta_i(\mathbf{r}-\mathbf{R}_i)^2}
\end{equation}
where $\zeta_i$ is an adjustable parameter related to the Gaussian width. 

The original model at zero applied potential can be rewritten as follows using the chemical potential equalization ansatz~\cite{York:1996ia} 
\begin{equation}
    U = U_0 + U_{q_0-\Delta\nu} + \frac{1}{2} \mathbf{c}^\top 
    \boldsymbol{\eta} \mathbf{c} +  \Delta \boldsymbol\nu^\top\mathbf{c} \label{tot_energy}
\end{equation}
where $U_0$ corresponds to the energy of electrode atoms in absence of an external potential (field). The term $U_{q_0-\Delta\nu}$ corresponds to the electrode-electrolyte interaction (so electrostatic interactions between the atomic charges of electrolyte atoms and the base charges $\mathbf{q}_0$ of electrode atoms plus their van der Waals interactions). $\boldsymbol\eta$ is the hardness kernel, describing the interaction between response charges and $\Delta \boldsymbol\nu$ is the potential generated by the electrolyte at the electrode atom sites. This energy is minimized with respect to the response charge $\mathbf{c}$ at each MD time-step under the constraint of charge neutrality.

With this model, a full-cell setup can be realized with two electrodes kept at constant potentials. If one electrode is set at zero potential we can enforce a constant potential difference $\Delta\psi$ between the electrodes by altering the model Hamiltonian as

\begin{equation}
    U_{\Delta \psi} = U - \sum_j \Delta\psi c_j,
\end{equation}

where the sum is over all charges belonging to one electrode. This requires an electrolyte centered cell and the use of 2D PBCs, which makes the potential calculation computationally more expensive than the Ewald summation with 3D PBCs.

Instead, with finite-field methods and 3D PBCs, the electrode charges are now coupled to an applied electric field $E$ through

\begin{equation}
    U_E = U - \Omega E P_z
\end{equation}

where $P_z = \frac{1}{\Omega}\mathbf R_z \cdot \mathbf c$ is the polarization in the direction of the electric field. In this scenario, one electrode with a single inner potential was employed instead and we can use an electrode centered cell. The cell potential bias now comes from the different charges at the two electrode surfaces, in analogy with the example in Figure \ref{Volta_CT}b. \added{Knowing the constant $E$ field directly leads to constant potential simulation under 3D PBCs, applying a potentiostat on top of constant $D$ simulations~\cite{Stengel:2009cd, Zhang:2016cl} may seem to be a detour. Nevertheless, interesting attempts have been made to control the potential difference using the constant electric displacement $D$ Hamiltonian, which generates a constant $\langle E \rangle$ ensemble instead~\cite{2021.Deissenbeck}. }

\section{Potential of metal versus potential of electrolyte}
So far, the focus of our discussions is on the potential control of metal. However, in an electrochemical cell with an active redox couple electrons can transfer between electrodes and solution. If we attempt to control the potential by just fixing the electrochemical potential of the metal electrode, $\tilde \mu_e^{\textrm{M}}$, there is no guarantee that the electrode potential is in equilibrium with the redox couple wihtin the time scale of simulation. For the chemical reaction 
\begin{equation}
    \text{O} + ne \rightleftharpoons \text{R}
\end{equation}
it is in equilibrium only if 
\begin{equation}
    \tilde \mu_{e}^{\textrm{M}} = \tilde \mu_{\textrm{R}}^\textrm{s}-\tilde \mu_{\textrm{O}}^{\textrm{s}}, 
\end{equation}
otherwise the system is under an overpotential. This means the electrified interface requires the counter-ions to form a concentration gradient in equilibrium. Therefore, to simulate an electrochemical cell under constant potential conditions, it is crucial to also take into account the potential of the electrolyte.

This is indeed very challenging as it requires a grand canonical ensemble for the ions and the corresponding equilibration times goes far beyond the reach of standard AIMD simulations.  As a matter of fact, a number of selected works listed in the Table 1 only used water instead of electrolyte solution when building a solid-liquid interfacial system. Therefore, strictly speaking, these setups simulated a \added{(nano)}capacitor rather than an electrochemical interface. 

Despite being challenging, potential control of the electrolyte can be achieved in various ways. With the help of liquid state theories,  a molecular description of the electrolyte can be introduced, e.g. in  joint DFT (JDFT)~\cite{jdft07, jdft12} where the solution is described by classical DFT~\cite{2022.Wuij}  or in conjunction with the reference interaction site method (RISM)~\cite{1972.Chandler, nishihara}. With these descriptions of the electrolyte, the effect of an ion reservoir can be achieved, comparable to the electron reservoir in Type I methods. In Type II \& III methods, explicit ions and electrolyte solution can be introduced  and one can explore the classical MD simulation to speed-up the equilibration of the electrolyte solution and bring counter-ions to their equilibrium positions next to the electrified interfaces. Similar to the case of metal in Type II \& III methods where two sides of electrode are under grand canonical condition but the whole piece of metal is still canonical, the same principle also applies to the electrolyte solution for its potential control.

\section{Summary and outlook}

In this short review, we have sorted computational methods for potential control in MD simulations of the metal-electrolyte interface into three categories. In general, the potential control can be achieved either through a grand canonical treatment of a half-cell model or a canonical treatment of the full-cell model. In both cases, charge transfer is induced by equalization of the Fermi level of a system by the exchange of electrons with an external reservoir or via redistribution of electrons within the system. 

 Work is still in progress on applying the finite-field methods to DFTMD simulations of the metal-electrolyte interface. In particular, it is interesting to see how the constant electric displacement method, that has been shown to be useful to compute Helmholtz capacitance of protonic double layers at metal oxide-electrolyte interface~\cite{zhang2019coupling, jia2021origin}, can be also applied to the metal-electrolyte interfaces. 

Finally, it is impossible not to mention machine-learning potentials (MLP) when discussing MD simulations nowadays. Indeed, a number of implementations toward the goal of modelling electrochemical systems using machine-learning accelerated atomistic simulations have emerged~\cite{dufils2023pinnwall, grisafi2023predicting}. Their further developments and the full harnessing of the scalability of MLPs are expected to play a crucial role in the future of  DFTMD simulations of electrified interfaces beyond slab geometry and short dynamics. 

\section*{Declarations of interest}
The authors declare that they have no known competing financial interests or personal relationships that could have appeared to influence the work reported in this paper.

\section*{Acknowledgement}

This project has received funding from the European Research Council (ERC) under the European Union's Horizon 2020 research and innovation programme (grant agreement No. 949012). We thank M. Sprik for critically reading the manuscript and helpful suggestions. 


 \bibliographystyle{elsarticle-num} 





\end{document}